\documentclass[letterpaper,10 pt,twocolumn,conference,numbers,sortcompress]{IEEEtran}
\IEEEoverridecommandlockouts
\usepackage{cite}
\usepackage{amsmath,amssymb,amsfonts}
\usepackage{algorithmic}
\usepackage{graphicx}
\usepackage{textcomp}
\usepackage{xcolor}
\usepackage{url}
\usepackage{hyperref}
\usepackage{blindtext}
\usepackage{caption}
\usepackage[utf8]{inputenc}
\usepackage{booktabs}
\usepackage{multirow}
\usepackage[section]{placeins}
\usepackage{caption}
\hypersetup{
    colorlinks = true,
    linkcolor = blue,
    urlcolor = blue,
}
\def\BibTeX{{\rm B\kern-.05em{\sc i\kern-.025em b}\kern-.08em
    T\kern-.1667em\lower.7ex\hbox{E}\kern-.125emX}}
\begin{document}

\title{\LARGE \bf{Analyzing Collision Rates in Large-Scale Mixed Traffic Control via Multi-Agent Reinforcement Learning \\}
{}
\thanks{}
}

\author{\IEEEauthorblockN{Muyang Fan}
\IEEEauthorblockA{\textit{Department of Computer Science} \\
\textit{University of Memphis}\\
Memphis, USA \\
} 

}

\maketitle

\begin{abstract}
Vehicle collisions remain a major challenge in large-scale mixed traffic systems, especially when human-driven vehicles (HVs) and robotic vehicles (RVs) interact under dynamic and uncertain conditions. Although Multi-Agent Reinforcement Learning (MARL) offers promising capabilities for traffic signal control, ensuring safety within such environments is still difficult. As a direct indicator of traffic risk, the collision rate must be understood and incorporated into traffic control design.

This study investigates the primary factors influencing collision rates in a MARL-governed Mixed Traffic Control (MTC) network. We examine three dimensions: total vehicle count, signalized–unsignalized intersection configurations, and turning-movement strategies. Through controlled simulation experiments, we evaluate how each factor affects collision likelihood.

Results show that the collision rates are susceptible to traffic density, signal coordination level, and turning-control design. The findings provide practical insights for improving the safety and robustness of MARL-based mixed traffic control systems, supporting the development of intelligent transportation systems where both efficiency and safety are jointly optimized.
\end{abstract}

\section{Introduction}
Vehicle collisions remain one of the most critical challenges in modern transportation systems, not only posing direct threats to human life and roadway safety but also reducing traffic efficiency and imposing substantial economic burdens. According to the previous annual report by the U.S. National Highway Traffic Safety Administration (NHTSA), traffic collisions in the United States result in 35,000 - 43,000 fatalities and more than 2.7 million injuries each year, making traffic crashes one of the most significant public health issues in the nation~\cite{NHTSA2023}. Economically, the U.S. Department of Transportation (DOT) reports that the annual economic loss caused by traffic collisions exceeds US\$1 000 billion, including medical expenses, productivity loss, property damage, law enforcement costs, and congestion-related delays~\cite{DOT2022}.

Although autonomous driving technologies and intelligent traffic networks have made significant progress in recent years, ensuring safety in large-scale \textit{Mixed Traffic Control} (MTC) systems via \textit{Multi-Agent Reinforcement Learning} (MARL) remains highly challenging.~\cite{Rakha2019} In mixed environments where human-driven vehicles (HVs) and robot vehicles (RVs) coexist, the stochastic nature of human driving behaviour, dynamic signal timing variations, and heterogeneous traffic densities collectively introduce unpredictable risks, thereby increasing uncertainty for Reinforcement Learning (RL) and multi-agent coordination systems~\cite{Shladover2018}.

Among various performance indicators, the vehicle \textit{collision rate} is one of the most direct and quantitative measures of traffic safety~\cite{He2020}. Understanding the key factors that influence collision rates not only helps identify potential safety risks but also provides essential insights for developing more adaptive and robust traffic control strategies. Incorporating collision rate into control objectives enables MARL-based traffic systems to better balance efficiency and safety, ensuring safer and more stable operation under complex and dynamic traffic conditions.

This study systematically analyzes the key factors that affect collision rates in a large-scale mixed traffic network governed by MARL. Specifically, we investigate three dimensions:
\begin{itemize}
    \item \textbf{Signalized--unsignalized configuration ratio}: representing different levels of Robotic vehicles control coordination;
     \item \textbf{Total number of vehicles}: reflecting the impact of traffic density on collision risk;
    \item \textbf{Turning-movement control}: changing left-turn into straight movements could reduce conflicts.
\end{itemize}

By conducting controlled experiments on a large-scale traffic network of 14 intersections using the Simulation of Urban MObility (SUMO)~\cite{lopez2018microscopic}under various traffic control conditions, this work aims to uncover the underlying mechanisms that lead to collision events and to provide quantitative guidance for designing safer large-scale mixed traffic control systems.
The code of our work is available at \url{https://github.com/cgchrfchscyrh/MixedTrafficControl_IROS}. 

\section{Related Work}

Recent studies have shown that traffic flow in mixed traffic environments exhibits increased unpredictability and instability~\cite{UAVPET}. Further research indicates that in high-density scenarios, interactions between human-driven vehicles (HVs) and robot vehicles (RVs) introduce nonlinear risks that substantially increase the probability of collisions~\cite{PETTTC}. However, most of the existing work is limited to small-scale corridors or isolated intersections, leaving a lack of systematic collision analysis at the scale of large traffic networks.

Recent advances demonstrate that Multi-Agent Reinforcement Learning (MARL), particularly in distributed forms, can enhance traffic throughput, reduce delays, and achieve scalable signal coordination in large urban networks~\cite{Marl1,Pan2025Review,Liu2025Large,Islam2025Heterogeneous,Fan2025OD,Wang2024Intersection,Wang2024Privacy,Poudel2024CARL,Poudel2024EnduRL,Villarreal2024Eco,Villarreal2023Pixel,Villarreal2023Chat
}. However, these studies predominantly focus on efficiency-related metrics (e.g., travel time, queue length), while largely neglecting safety considerations. Only a few attempts have incorporated risk-sensitive objectives into MARL, yet very limited work has explicitly adopted collision rate as a safety evaluation metric or examined it in mixed HV--RV environments.

Existing research also highlights that turning movements at intersections - especially left-turn versus opposing-through conflicts-are among the primary sources of angle collisions and severe crashes~\cite{leftturn1}. Previous studies have explored approaches such as geometric redesign, protected left-turn phases, and signal phase restructuring to reduce conflict points~\cite{leftturn2}. Although these methods have demonstrated effectiveness in low- to medium-volume scenarios, they have not been thoroughly validated in large-scale MARL-controlled mixed traffic networks, where the complexity of HV-RV interactions may introduce additional safety challenges.

Unlike recent and existing studies, which primarily focus on efficiency improvements in small-scale networks or single-intersection settings, our work performs a systematic, network-level analysis of collision rates in a 14-intersections mixed traffic system governed by MARL. Specifically, we examine how three conventional structural factors—the proportion of signalized versus unsignalized intersections, the overall traffic demand, and the elimination of left-turn movements—collectively influence collision outcomes in HV–RV environments, which represents one of the large-scale studies to explicitly evaluate collision rate as a primary safety metric under MARL control.

\section{Methodology}
The problem for robotic vehicles (RVs) operating at unsignalized intersections is modelled as a decentralized partially observable Markov decision process (Dec-POMDP), formally defined as a six-tuple $\mathcal{M} = \langle \mathcal{S}, \mathcal{U}, P, R, \mathcal{O}, \lambda \rangle$. Here, $\mathcal{S}$ denotes the global state space of the mixed-traffic environment, while $\mathcal{U}$ represents the joint action space comprising individual maneuver decisions made by RVs. The stochastic transition function $P(\mathbf{s}_{t+1} \mid \mathbf{s}_t, \mathcal{U})$ reflects the evolution of the state of the system driven by vehicle interactions, and the reward function $R(\mathbf{s}_t, \mathcal{U})$ evaluates immediate performance in terms of safety and mobility. Each RV receives a private observation $\mathbf{s}_{t}^{(j)} \in \mathcal{O}$ rather than full access to the global state, which enforces decision-making under partial observability. The discount factor $\lambda \in (0,1)$ regulates the contribution of long-term effects. This formulation captures decentralized decision-making under incomplete information, which is
fundamental for cooperative control in large-scale mixed-traffic systems~\cite{Jiang2022UniComm,Kolat2023MARTSC}.

\subsection{Reinforcement Learning Formulation}
Rainbow DQN is adopted in this study due to its stability, sample efficiency, and robustness in complex mixed traffic environments. Traditional DQN often suffers from Q-value overestimation, unstable learning, and slow convergence under the high stochasticity and partial observability inherent in MARL-based traffic networks. Rainbow integrates several key enhancements--including Double Q-learning, Dueling Networks, Prioritized Replay, and Distributional RL (C51), which together provide more accurate value estimation, improved training stability, and better sensitivity to sparse high-risk events such as collisions~\cite{hessel2018rainbow,wang2016dueling,bellemare2017distributional}. The RVs are trained using a multi-agent variant of the Rainbow Deep Q-Network (Rainbow-DQN) through the integration of distributional reinforcement learning, Double Q-learning, and Dueling Network structures.
During training, the Q-function is modeled
using a Dueling Network architecture to separately represent the state-dependent utility and the relative advantage of
each action. Formally, the approximation is expressed as:
\begin{equation}
Q_{\varphi}(\mathbf{s},u)
=
U_{\varphi}(\mathbf{s})
+
\Biggl[
A_{\varphi}(\mathbf{s},u)
-
\frac{1}{|\mathcal{U}|}
\sum_{u' \in \mathcal{U}}
A_{\varphi}(\mathbf{s},u')
\Biggr],
\label{eq:exp_dueling_rewrite}
\end{equation}
where $U_{\varphi}(\mathbf{s})$ denotes the estimated baseline value of observation $\mathbf{s}$ and
$A_{\varphi}(\mathbf{s},u)$ quantifies the relative preference for selecting action $u$. The subtractive
normalization ensures that the advantage function remains identifiable and stabilizes optimization in
multi-agent settings.

Each RV $j \in \mathcal{M}$ optimizes a shared stochastic driving policy $\pi_{\vartheta}$ by maximizing the 
discounted sum of future rewards:
\begin{equation}
J(\pi_{\vartheta})
=
\mathbb{E}_{\pi_{\vartheta}}
\left[
\sum_{t=0}^{T-1}
\lambda^{t}
R_{t}^{(j)}
\right].
\label{eq:method_obj}
\end{equation}

Its observation vector aggregates lane-level traffic information:
\begin{equation}
\mathbf{s}_{t}^{(j)}
=
\left\{
L_{k},\ d_{k},\ \eta_{k}
\ \middle|\ 
k \in \mathcal{L}
\right\},
\label{eq:method_obs}
\end{equation}
where $L_{k}$ is the queue length, $d_{k}$ the delay associated with lane $k$, and $\eta_{k}$ an indicator of
whether the conflict zone is currently occupied.

The RV selects from two manoeuvre actions:
\begin{equation}
u_{t}^{(j)}
\in
\left\{
u^{\mathrm{go}},\,
u^{\mathrm{stop}}
\right\}.
\label{eq:method_action}
\end{equation}

The reward function balances mobility and safety using a weighted structure:
\begin{equation}
R_{t}^{(j)}
=
\alpha R_{\mathrm{flow}}
+
\beta R_{\mathrm{safety}},
\qquad
\alpha>0,\ \beta<0,
\label{eq:method_reward}
\end{equation}

where
\begin{equation}
R_{\mathrm{flow}}
=
\begin{cases}
+d_{k}, & u_{t}^{(j)} = u^{\mathrm{go}},\\[3pt]
-d_{k}, & u_{t}^{(j)} = u^{\mathrm{stop}},
\end{cases}
\quad
R_{\mathrm{safety}}
=
-
\mathbb{I}(\text{collision}).
\label{eq:method_reward_terms}
\end{equation}

\subsection{Modeling of Human-driven Vehicles}
All human-driven vehicles (HVs) are governed by the Intelligent Driver Model (IDM)~\cite{treiber2000congested} to ensure realistic longitudinal
motion based on dynamic headways and relative velocities. RVs adopt IDM behavior outside a predefined safety zone 
and activate the learned policy within a $30$~m radius around intersections, allowing policy transition.

\subsection{Integrated Mixed Control of Intersections}
The traffic network incorporates signalized and unsignalized intersections. Signalized intersections are governed by traffic lights, whereas unsignalized intersections rely on RVs executing the learned Stop/Go policy. This integrated structure forms a scalable mixed control architecture that mirrors real-world urban traffic, allowing coordinated decision-making between RL-controlled RVs and conventional signal control~\cite{Peng2021CAVUnsig,Shi2022HybridUnsigRL,Spatharis2024MARLUnsignalized}.

\subsection{Evaluation Metrics}
We use \textit{Collision Rate (CR)} as the primary safety metric for evaluating mixed traffic performance~\cite{Li2021}. Collision Rate measures the proportion of vehicles that experienced at least one collision during the simulation.
Collision Rate is defined as:
\begin{equation}
CR = \frac{N_{\text{collided}}}{N_{\text{departed}}}
\label{eq:cr_overall}
\end{equation}

where:
\begin{itemize}
    \item $N_{\text{collided}}$ is the total number of vehicles involved in one or more collision events,
    \item $N_{\text{departed}}$ is the total number of vehicles that successfully departed from the network.
\end{itemize}
\section {Experiments and Results}
\subsection{Exeperiment Set-up}
This study evaluates the MARL-based mixed traffic control framework on a 
14-intersection urban network consisting of both signalized and unsignalized 
intersections under different scenarios. In this work, we denote 
\textbf{U} as an \textit{unsignalized intersection} governed by the RV-based 
MARL controller, and \textbf{S} as a \textit{signalized intersection} operated by a traditional fixed-time signal controller. A notation of the form \textit{U + S} represents the total network composition in terms of the number of unsignalized (U) and signalized (S) intersections. To investigate how network-level control structures influence collision dynamics, we construct five configurations with different proportions of signalized and unsignalized intersections:

\begin{itemize}
    \item\textit{ 12 U + 2 S} intersections,
    \item \textit{10 U + 4 S }intersections,
    \item \textit{8 U + 6 S }intersections,
    \item \textit{6 U + 8 S }intersections,
    \item \textit{4 U + 10 S }intersections.
\end{itemize}

The learning architecture adopts three fully connected layers with 512 neurons each, while employing a prioritized experience replay with capacity 50,000 ($\alpha = 0.5$) to accelerate convergence toward optimal policies. Key hyperparameters include discount factor $\gamma = 0.99$, learning rate $5\!\times\!10^{-4}$, mini-batch size of 32, and a control zone radius of 30 meters, determining when RVs transition from human-behavior emulation (IDM) to learned policies. Performance evaluation is conducted over 100 simulation rollouts, each lasting 1,000 s with approximately 8,000 and 6000 vehicles, separately, circulating in the network, ensuring statistical robustness.

To assess scalability and adaptability under increasing automation, four RV penetration levels (25\%, 40\%, 60\%, and 80\%) are examined. All RVs share the same decentralized policy and act independently based on their local observation
\(\mathbf{s}_{t}^{(j)}\), as presented in Methodology. Training is performed for 1,000 episodes for every experimental setting utilizing the Ray multi-agent learning framework.

The experimental study consists of three major components, each designed to 
examine a different factor influencing collision rates in mixed traffic 
control (MTC) environment under MARL-based decision making:

\begin{itemize}
    \item[a] \textbf{Impact of Signalized/Unsignalized Intersection Configurations:}  
    We evaluate a 14-intersections urban network under five mixed-control 
    configurations: 
    \textit{12 U + 2 S}, 
    \textit{10 U + 4 S}, 
    \textit{8 U + 6 S}, 
    \textit{6 U + 8 S}, 
    and \textit{4 U + 10 S}.  
    These configurations are compared against a baseline to analyze how the proportion of traditional signalized intersections versus RV-controlled unsignalized intersections affect overall collision rates, and whether introducing RV-controlled intersections provides measurable safety benefits in MTC via MARL like conventional method does~\cite{Li2024MixedSignalized}. Signalized and unsignalized intersections rely on fundamentally different conflict–resolution mechanisms: signalized intersections separate movements temporally through phase scheduling, whereas unsignalized intersections resolve conflicts spatially through vehicle interactions. Varying the signalized--unsignalized composition therefore enables a systematic assessment of how intersection control structure influences collision dynamics.

    \item[b] \textbf{Impact of Traffic Demand Variation:}  
    To investigate the influence of traffic density, we reduce the total number of vehicles in each configuration from 7897 to 5911. Traffic demand directly modulates network density and the frequency of potential conflict events, making it one of the primary amplifiers of collision risk. In MARL-controlled systems, traffic density additionally affects the stability and sensitivity of learned policies due to dynamic state coupling and partial observability. This allows us to assess whether changes in traffic demand alter collision patterns in an MTC-via-MARL system and whether such effects align with those commonly observed in conventional traffic networks.

    \item[c] \textbf{Impact of Turning-Movement Modification:}  
    All left-turn movements in the five configurations are replaced with 
    straight-through movements while keeping traffic volume unchanged.  
    Intersection directional modification-- particularly on left-turn is well known as a major contributor to reduce angle collisions in traditional traffic engineering~\cite{Liang2017,FHWA2020}. Eliminating left-turn movements provides a mechanism to examine whether classical safety-enhancing strategies remain effective and the sensitivity of MARL-based mixed traffic control to directional movement changes.

\end{itemize}

\subsection{Results}
\begin{figure*}
    \centering
    \includegraphics[width=0.99\linewidth]{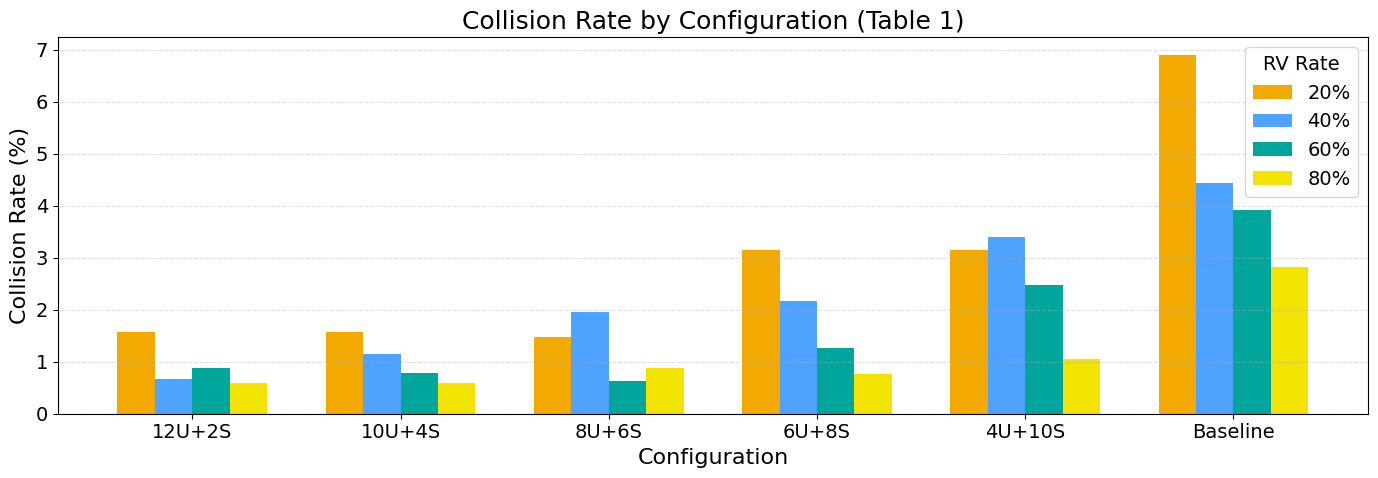}
    \caption{\small{The collision rate under different configurations: Collision rates under different signal and unsignal intersection configurations. As the number of RV-controlled unsignal intersections increases, the overall collision rate exhibits a decreasing trend. This indicates that RV-based control strategies are more effective than traditional traffic signal control in enhancing safety performance within large-scale mixed traffic networks.} }
    \label{fig1}     
\vspace{-1.5em}
\end{figure*}

\begin{figure*}
    \centering
    \includegraphics[width=0.99\linewidth]{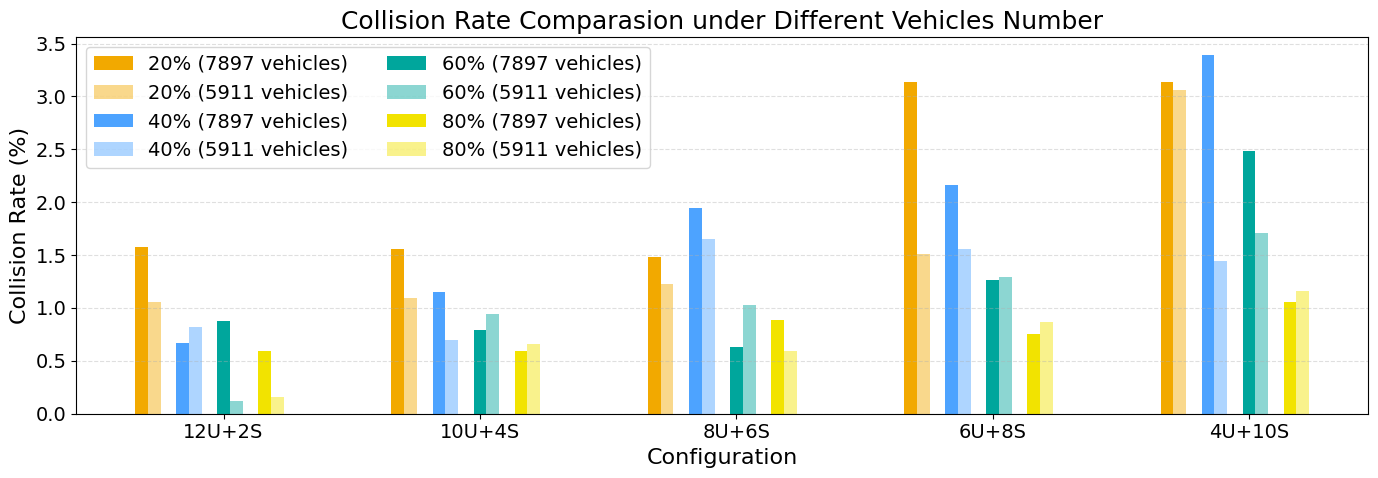}
   
    \caption{\small{The collision rate under different vehicle numbers: Reducing the overall vehicle volume leads to a noticeable decrease in collision 
rates in large-scale mixed traffic control (MTC) networks managed by MARL. 
This reduction effect is more significant in configurations with fewer 
RV-controlled unsignal intersections.}}
    \label{fig2}     
\vspace{-1.5em}
\end{figure*}

\begin{figure*}
    \centering
    \includegraphics[width=0.99\linewidth]{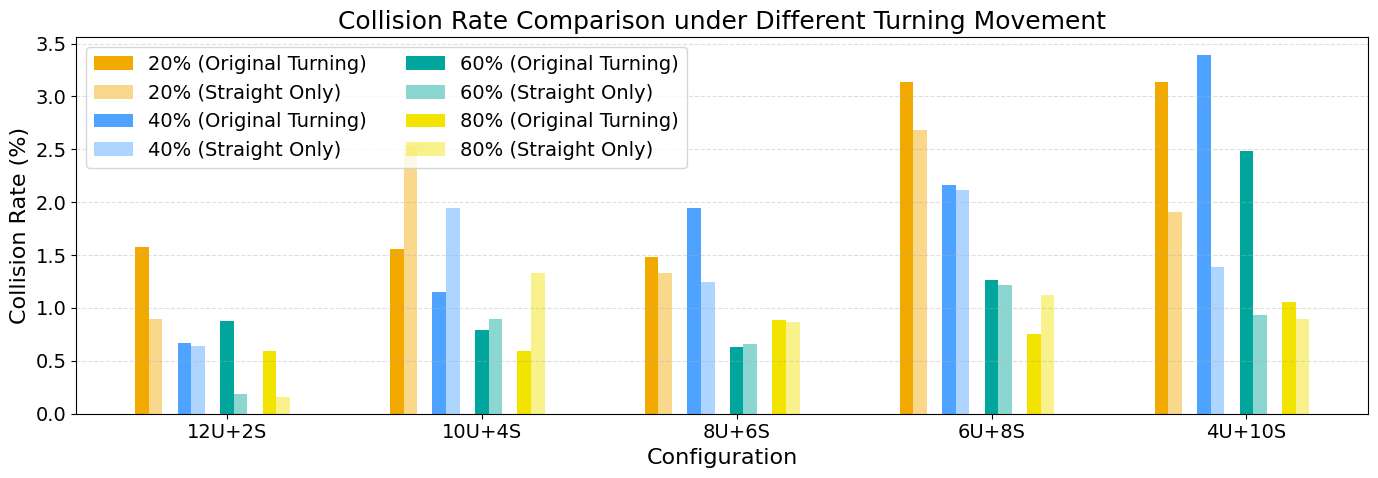}
    \caption{\small{The collision rate under different turning movements: With all left-turn movements converted into straight-through movements, 
most configurations exhibit a noticeable decrease in collision rates, 
indicating improved safety performance. However, the \textit{10U+4S} configuration shows an opposite trend, where collision rates increase across multiple RV penetration levels, suggesting that the removal of left-turn phases may cause increased conflict interaction under this specific network topology.} }
    \label{fig3}     
\vspace{-1.5em}
\end{figure*}

The experimental results presented in Fig.~\ref{fig1}, Fig.~\ref{fig2}, Fig.~\ref{fig3} and Table~\ref{tab1}, Table~\ref{tab2}, Table~\ref{tab3} demonstrate that increasing the proportion of unsignalized intersections, reducing overall traffic demand, and converting left-turn movements into straight-through flows generally lead to lower collision rates across the evaluated mixed traffic scenarios.

\begin{table}[htb] 
\centering
\footnotesize
\begin{tabular}{c|ccccc|c}
\hline
\multicolumn{7}{c}{\textbf{Collision Rate(\%)}} \\
\hline
\textbf{RV Rate} &
\textbf{12U+2S} &
\textbf{10U+4S} &
\textbf{8U+6S} &
\textbf{6U+8S} &
\textbf{4U+10S} &
\textbf{Baseline} \\
\hline
20\% & 1.580 & 1.562 & 1.484 & 3.138 & 3.140 & 6.890 \\
40\% & 0.667 & 1.148 & 1.950 & 2.162 & 3.390 & 4.430 \\
60\% & 0.879 & 0.788 & 0.634 & 1.266 & 2.480 & 3.920 \\
80\% & 0.595 & 0.593 & 0.884 & 0.754 & 1.060 & 2.820 \\
\hline
\end{tabular}
\captionsetup{justification=justified, singlelinecheck=false}
\caption{\small{Collision rates under five intersection configurations and the baseline scenario across different RV penetration rates, under a total of 7897 vehicles. Each configuration varies the proportion of unsignalized intersections controlled by robotic vehicles, enabling examination of how localized autonomy influences system-level safety. Generally, increasing RVs controlled unsignalized intersections reduced collision rates for the network of MTC via MARL.}}
\label{tab1}
\end{table}

\begin{table}[htb]
\centering
\footnotesize
\begin{tabular}{c|ccccc}
\hline
\multicolumn{6}{c}{\textbf{Collision Rate(\%)}} \\
\hline
\textbf{RV Rate} &
\textbf{12U+2S} &
\textbf{10U+4S} &
\textbf{8U+6S} &
\textbf{6U+8S} &
\textbf{4U+10S} \\
\hline
20\% & 1.060 & 1.096 & 1.230 & 1.510 & 3.060 \\
40\% & 0.823 & 0.7012 & 1.650 & 1.560 & 1.440 \\
60\% & 0.118 & 0.945 & 1.030 & 1.290 & 1.705 \\
80\% & 0.156 & 0.664 & 0.594 & 0.865 & 1.160 \\
\hline
\end{tabular}
\captionsetup{justification=justified, singlelinecheck=false}
\caption{\small{Collision rates under five intersection configurations across different RV penetration rates of total vehicles 5911. This highlights the sensitivity of mixed-traffic safety outcomes to network loading and demonstrates how lower demand interacts with autonomous negotiation. Generally, reducing the total amount vehicles in the network provides lower collision rates for the network of MTC via MARL.} }
\label{tab2}
\end{table}

\begin{table}[htb]
\centering
\footnotesize
\begin{tabular}{c|ccccc}
\hline
\multicolumn{6}{c}{\textbf{Collision Rate(\%)}} \\
\hline
\textbf{RV Rate} &
\textbf{12U+2S} &
\textbf{10U+4S} &
\textbf{8U+6S} &
\textbf{6U+8S} &
\textbf{4U+10S} \\
\hline
20\% & 0.898 & 2.540 & 1.331 & 2.680 & 1.910 \\
40\% & 0.6405 & 1.950 & 1.245 & 2.1112 & 1.390 \\
60\% & 0.185 & 0.900 & 0.663 & 1.220 & 0.930 \\
80\% & 0.154 & 1.330 & 0.870 & 1.120 & 0.900 \\
\hline
\end{tabular}
\captionsetup{justification=justified, singlelinecheck=false}
\caption{\small{Collision rates under five intersection configurations across different RV penetration rates under circumstances of converting left turns to straight, under a total of 7897 vehicles. The results illustrate how geometric and operational alterations interact with autonomous control across different mixed-intersection layouts. Generallyalthough it does not show effectiveness in the 4S+10u configuration, converting left turns to straight lines provides lower collision rates for the MTC network via MARL. }}
\label{tab3}
\end{table}

\
The  details of the experimental results  are presented as follows:
\begin{itemize}
    \item a) \textbf{Effect of Signalized--Unsignalized Configurations:}
    Across the five configurations with different proportions of signalized and unsignalized intersections, a consistent downward trend in collision rates is observed as the number of RV-controlled unsignalized intersections increases. The \textit{2 signalized + 12 unsignalized} network achieves the lowest overall collision levels, while the minimum collision rate (0.595\%) occurs under the \textit{4 signalized + 10 unsignalized} configuration at an RV penetration of 80\%. These findings demonstrate that unsignalized intersections controlled by RVs offer substantially improved safety performance relative to conventional signalized intersections. Increasing the share of RV-controlled intersections effectively reduces conflict occurrences and mitigates safety risks in MARL-driven mixed traffic control (MTC) networks.

    \item b) \textbf{Effect of Reducing Network Traffic Demand:}
    When the total inflow to the traffic network is reduced, the collision rates across the original five configurations generally decrease. However, this mitigation effect becomes less consistent at RV penetration levels of 60\% and 80\%, where certain configurations show unexpected increases. Under reduced demand, the lowest collision rate is found in the \textit{2 signalized + 12 unsignalized} configuration at an RV rate of 60\% (0.118\%). In the \textit{4 unsignalized + 10 signalized} configuration, the greatest improvement occurs at an RV rate of 40\%, producing a 57.5\% reduction in collision rate. This demonstrates that reducing overall network inflow can alleviate collision risks, although its benefits vary across configurations and penetration levels, indicating that moderating overall traffic demand can reduce collision risks in MARL-controlled networks.
    
    \item c) \textbf{Effect of Converting Left-Turn Movements to Straight Movements:} 
    After converting all left-turn movements in the network to straight-through movements, the collision rates of the original configurations generally decline. However, this modification results in an overall increase for the \textit{4 signalized + 10 unsignalized} configuration, indicating limited effectiveness in this structure. The lowest collision rate under this modification (0.154\%) is achieved in the \textit{2 signalized + 12 unsignalized} configuration at an RV rate of 80\%. For the \textit{4 unsignalized + 10 signalized} configuration, the largest reduction occurs at an RV rate of 60\%, achieving a 62.3\% decrease. These findings indicate that converting left-turn movements to straight-through flows can reduce conflict points and improve safety in MARL-governed mixed traffic systems, but the magnitude of improvement might be dependent on the geographic distribution of signalized versus unsignalized intersections.
\end{itemize}

\section{Conclusion and future work}
This study shows that intersection control structure, traffic demand intensity, and turning-movement design each have a substantial impact on collision rates in MARL-controlled mixed traffic networks. Increasing the share of RV-controlled unsignalized intersections consistently reduces collision rates, underscoring the safety benefits of decentralized RV decision-making. Reducing network inflow further lowers collision rates in most configurations, though this effect becomes less stable at medium–high RV penetration levels, indicating a non-linear interaction between traffic density and MARL policies. Eliminating left-turn movements generally improves safety but produces adverse effects in some configurations, suggesting that its effectiveness depends on network geometry and control structure.

In general, the results reveal that classical safety improvement strategies—expanding unsignalized RV control, modifying traffic demand, and reducing conflict points—remain effective under mixed autonomy driven by MARL. The study fills a key gap by providing network-level, collision-focused insights and offers theoretical guidance for designing safer and more robust MARL-based traffic control systems.

Due to time limitations, the present work does not examine a broader range of signalized--unsignalized configurations, and all intersections are modeled as standard layouts without roundabouts. Additionally, the network traffic demand follows a single dominant directional pattern. Future work will extend the experiment design to incorporate richer combinations of control structures, more diverse intersection geometries, and multiple traffic demand distributions to further enhance the generalizability of the conclusions.

Many future research directions exist. First, we aim to evaluate our method in more sophisticated and expansive scenarios that incorporate mobile data calibration to accurately represent real-world traffic patterns~\cite{Guo2024Simulation,Chao2020Survey,Li2017CityFlowRecon,Wilkie2015Virtual}. Second, we plan to enhance our framework by integrating additional general information, including traffic state forecasts and vehicle trajectory data, which may yield further improvements~\cite{Poudel2025Urban,Raskoti2025MIAT,Lin2022GCGRNN,Lin2019BikeTRB,Poudel2021Attack,Lin2019Compress,Li2018CityEstIET,Li2017CitySparseITSM
}. Third, we seek to strengthen the robustness of our approach by accounting for potential adversarial attacks. Finally, we would like to explore the integration of this work into the testing of autonomous driving technology~\cite{Villarreal2024AutoJoin,Lin2022Attention,Poudel2022Micro,Shen2021Corruption,Shen2022IRL,Li2019ADAPS}.

\bibliographystyle{IEEEtran}
\bibliography{ref}

\end{document}